\documentclass{elsart4}

\usepackage{natbib}

\usepackage{amssymb}
\usepackage{bm}
\usepackage{graphicx}

\begin{document}

\begin{frontmatter}



\title{Large-scale electronic-structure theory and 
nanoscale defects formed in cleavage process of silicon}


\author{T. Hoshi\corauthref{cor1}\thanksref{label1}\thanksref{label2}}
 \ead{hoshi@coral.t.u-tokyo.ac.jp}
 \ead[url]{http://fujimac.t.u-tokyo.ac.jp/lses/}
\author{R. Takayama\thanksref{label3}\corauthref{label4}}
\author{Y. Iguchi \thanksref{label1}}
\author{T. Fujiwara\thanksref{label1}\thanksref{label2}}

\corauth[cor1]{corresponding author \\
\quad Takeo Hoshi \\
\quad University of Tokyo \\
\quad 7-3-1 Hongo, Bunkyo-ku, Tokyo, Japan\\
\quad tel +81-3-5841-6812; \ fax +81-3-5841-6869}

\address[label1]{Department of Applied Physics, University of Tokyo, Tokyo, Japan}
\address[label2]{Core Research for Evolutional Science and Technology (CREST), 
Japan Science and Technology Agency, Saitama, Japan.}
\address[label3]{Research and Development for Applying Advanced Computational Science and Technology (ACT-JST),  
Japan Science and Technology Agency, Saitama, Japan.}

 \thanks[label4]{Present address:Canon Inc., Analysis technology center, 
30-2 Shimomaruko 3-Chome, Ohta-ku, Tokyo 146-8501, Japan}


\begin{abstract}
Several methods are constructed for large-scale electronic structure calculations.
Test calculations are carried out with up to $10^7$ atoms. 
As an application, cleavage process of  silicon is investigated 
by molecular dynamics simulation with 10-nm-scale systems.  
As well as the elementary formation process of 
the $(111)$-($2 \times 1$) surface, 
we obtain nanoscale defects, 
that is, step formation and bending of cleavage path 
into favorite (experimentally observed) planes. 
These results are consistent to experiments. 
Moreover, the simulation result 
predicts an explicit step structure on the cleaved surface, 
which shows a bias-dependent STM image. 
\end{abstract}

\begin{keyword}
order-$N$ electronic-structure theory, 
nanoscale defect, 
surface process of silicon, 
bias-dependent STM image.
\end{keyword}

\end{frontmatter}

\pagebreak

\section{Introduction \label{INTRO}}

Quantum mechanical 
(electronic structure) calculation with 10-nm-scale systems 
is of great importance in the present semiconductor technology. 
Particularly,  
dynamical simulations are highly desirable 
so as to explore industrial processes. 
These simulations, however, are  
quite difficult for the present standard methodology,
such as the Car-Parrinello method \cite{CP},
owing to its heavy computational cost.
So as to overcome the difficulty, 
methodology for large systems, with thousands of atoms or more, 
has been focused from 1990's.  \cite{KOHN96, REVIEW-ON}
Its mathematical foundation is the calculation of one-body density matrix,  
instead of one-electron eigenstates. 
In this paper, we will show the investigation of nanoscale defects 
 formed in cleavage process of silicon
\cite{HOSHI2004A},
which is an example of our newly-developed practical methods 
of large-scale electronic structure calculations. 
\cite{HOSHI2000A,HOSHI2001A,HOSHI2003A,GESHI,TAKAYAMA2004A, 
TAKAYAMA2004B}
These methods are general and are applicable 
to not only single-atom defects but also nanoscale defects.

\begin{figure}[tbh]
\begin{center}
  \includegraphics[width=7cm]{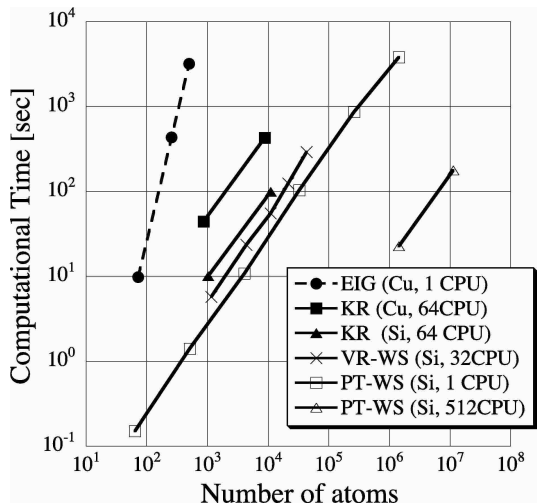}
\end{center}
\caption{
The computational time for bulk Si or Cu as a function 
of the number of atoms ($N$),
up to 11,315,021 atoms 
(Refs. \cite{HOSHI2003A, HOSHI2004A} and this work).
The time was measured  
for electronic structure calculation 
with a given atomic structure.
The calculations were carried out
by the conventional eigenstate calculation (EIG) 
and by our methods for large system; 
(i) Krylov-subspace method (KR) 
with subspace-diagonalization procedure, 
(ii) variational Wannier-state method (VR-WS)
and (iii) perturbative Wannier-state method (PT-WS). 
For \lq 1CPU' computations, 
we used single Pentium 4$^{\rm TM}$ processor in 2 GHz. 
Parallel computations were carried out 
by SGI Origin 3800$^{\rm TM}$ (for PT-WS method), 
Origin 2800$^{\rm TM}$ (for VR-WS method) and
Altix 3700$^{\rm TM}$  (for KR method). 
}
\label{FIG-CPU-ORDER-N}
\end{figure}%

\section{Theory \label{THEORY}}

The  foundation of the large-scale electronic structure calculation
was established by W. Kohn \cite{KOHN96}. 
Any physical quantity $X$ can be given 
by the one-body density matrix $\rho$ as 
\begin{equation}
\langle \hat{X} \rangle 
 = {\rm Tr}[ \hat{\rho} \hat{X} ]
  = \int\int d{\bm r}d{\bm r}^\prime \rho ({\bm r},{\bm r}^\prime ) X({\bm r}^\prime, {\bm r}) ,
\label{eq:physica-quantity}
\end{equation}
where the density matrix is defined, from occupied one-electron eigenstates ${\phi_k(\bm{r})}$, as 
\begin{equation}
\ \hat{\rho} = \sum_k^{\rm occ.} | \phi_k \rangle \langle \phi_k  | .
\label{DM-DEF}
\end{equation}
The most important fact is that, 
in case that $\hat{X}$ is a  short-range  operator,
the physical quantity $\langle \hat{X} \rangle$ is contributed
only by the short-range component 
of the density matrix, 
even though the density matrix $\rho ({\bm r},{\bm r}^\prime )$  is of long range. 
Owing to this fact, 
W. Kohn proved that the density-functional ({\it ab initio}) theory 
can be constructed from the short-range component of the density matrix,
instead of eigenstates (Kohn-Sham orbitals).

Another methodological foundation for large-scale calculation
is the transferable Hamiltonians 
in the Slater-Koster (tight-binding) form. 
They are of short range and 
applicable to various circumstances,
{\it e.g.}, crystals, defects, liquid and surfaces.
Their success 
 has been known for decades 
and can be founded by the {\it ab initio} theory. 
\cite{NOTE-HOSHI2004A-2}

For practical large-scale calculations, 
we have developed a set of theories and program codes
with obtaining  density matrix. 
They are founded by 
generalized Wannier state 
\cite{HOSHI2000A,HOSHI2001A,HOSHI2003A,HOSHI2004A} 
or  Krylov subspace
(Refs. \cite{KRYLOV, TAKAYAMA2004A, TAKAYAMA2004B} and this work).
These methods are calculation methods for short-range component 
of the density matrix with a given Hamiltonian.
A bench mark is shown in Fig.~\ref{FIG-CPU-ORDER-N}, 
in which 
the computational time of our methods, 
unlike that of the conventional eigenstate calculation,  is
\lq order-$N$', or linearly proportional to the system size ($N$). 
Here and hereafter, 
all the calculations of Si are carried out 
using a typical transferable Hamiltonian. \cite{KWON} 
For the present calculation of Cu,
we used the Slater-Koster type Hamiltonian 
constructed from in  the LMTO theory \cite{LMTO}. 

As well as the calculation methods for density matrix, 
we also constructed a method for
Green function of large systems, 
which is based on Krylov subspace.
\cite{TAKAYAMA2004B,TAKAYAMA2004B-CODE}

\begin{figure}[tbh]
\begin{center}
  \includegraphics[width=7cm]{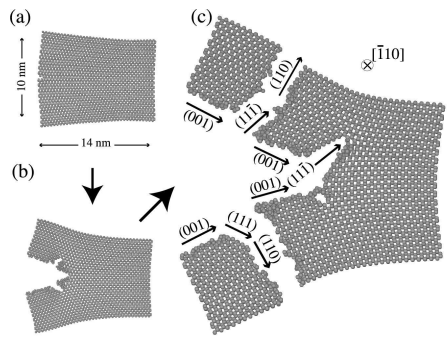}
\end{center}
\caption{
Bending of cleavage path into the favorite (experimentally observed) planes, 
$(111)$ and $(110)$ planes. 
\cite{HOSHI2004A}
The  time interval $\Delta t$  between the snapshots (a) and (b)
and that between (b) and (c) are $\Delta t =1$ ps and 2 ps, respectively.
}
\label{fig-frac-bend-awaji}
\end{figure}%

\section{Application to cleavage process in silicon} 

As the first application of our method, 
we focused on the cleavage process of silicon.
Molecular dynamics simulations were carried out 
with 10-nm scale samples or with upto $10^5$ atoms.
 \cite{HOSHI2003A,HOSHI2004A}

Since cleavage is a nonequilibrium process 
with a typical velocity scale of propagation velocity, 
its dynamical mechanism is essential. 
\cite{MOTT, BRITTLE}
Although cleavage process 
was simulated with electronic structure 
thus far, \cite{SPENCE93,GUMBSCH} 
its investigation was quite limited,
owing to its small system size of $10^2$ atoms. 
The present simulation method enables 
investigation beyond the above limitation.

\begin{figure}[thb]
\begin{center}
  \includegraphics[width=7cm]{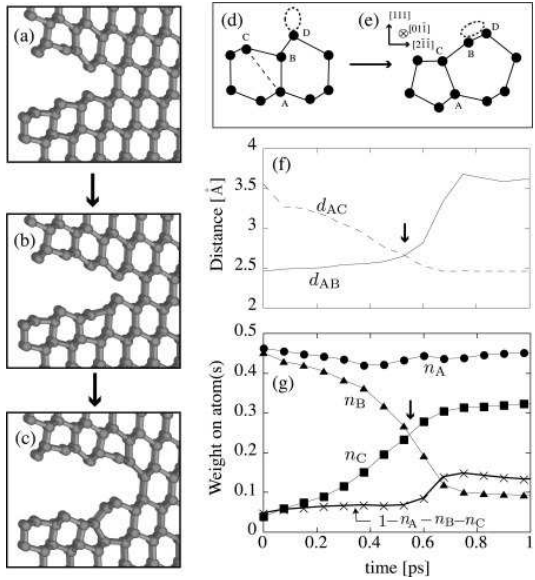}
\end{center}
\caption{
Elementary surface reconstruction process 
in silicon cleavage with $(111)-(2 \! \times \! 1)$ surface.
\cite{HOSHI2004A}
(a)-(c) : Snapshots of process. 
(d)-(e) : Schematic figures of 
(d) the buckled $(2 \! \times \! 1)$ structure 
that appears in the snapshot (b) 
and (e) the $\pi$-bonded (Pandey) $(2 \! \times \! 1)$ structure 
that appears in the snapshot (c).   
 The oval indicates the presence of a doubly occupied surface state. 
(e)-(f):  Quantum
mechanical analysis of a process in which the bonding 
wavefunction ($\phi_i$) between $A$ and $B$ sites 
changes into that between $A$ and $C$ sites
($A \! -\! B \, \Rightarrow A \! - \! C$);
Figure (f) shows the interatomic distances $d_{AB}$ and $d_{AC}$.
Figure (g) shows the occupation weight on 
the $A$, $B$ and $C$ atom sites ($n_A, n_B, n_C$) 
for the wavefunction $\phi_i$.  
A typical transitional state,
 with $d_{AB} \approx d_{AC}$  or $n_B \approx n_C$,
 appears at the time marked by arrows 
 in (f) and (g). 
}
\label{fig-hane-pan-fin}
\end{figure}

Numerical accuracy of the present simulation was confirmed, 
not only in elastic constants and surface energy, but also 
in the following quantities that can be measured by cleavage experiments;
(i) The critical stress intensity factor for cleavage 
$K_{\rm c}$ was evaluated to be
$K_{\rm c} = 0.7 {\rm MPa \sqrt{m}}$, 
which agrees well with experimental values of
$K_{\rm c} = 0.65 - 1.24 {\rm MPa \sqrt{m}}$ .~\cite{SPENCE93}
(ii) The cleavage propagation velocity  was evaluated to be
$v  \approx 2$nm/ps=2km/s,
which agrees well with a recent experimental value of 
$v = 2.3 \pm 0.3$km/s. 
\cite{HAUCH1999}

\subsection{Bending of cleavage path into favorite planes}

Experimentally, 
the cleavage plane of silicon is 
the (111) plane with 
a metastable $(2 \times 1)$  reconstruction
or the (110) plane (less favorable). 
\cite{NOTE-HOSHI2004A}
The appearance of these cleaved surfaces 
cannot be explained from  the traditional approach with surface energy. 
As a fruitful result of the present 10-nm-scale simulation, 
the experimental preference of 
cleaved surfaces is reproduced by bending of cleavage path; 
As a typical result,
Fig.~\ref{fig-frac-bend-awaji} shows that,  
even if the cleavage initiates artificially on another plane
($(001)$ plane), 
the cleavage path is bent into 
the experimentally observed planes ($(111)$ and $(110)$ planes). 
Moreover, we obtained a well-defined surface reconstruction on 
the $(111)$ and $(110)$ planes after bending \cite{HOSHI2004A}.

\begin{figure}[thb]
\begin{center}
  \includegraphics[width=7cm]{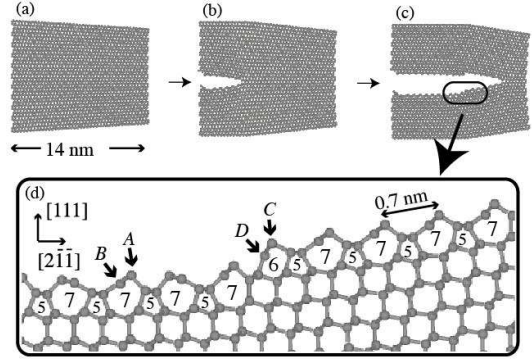}
\end{center}
\caption{
Stable (experimentally observed) cleavage mode 
on $(111)-(2 \! \times \! 1)$ surface. 
\cite{HOSHI2004A}
(a)-(c):Successive snapshots with a time interval of
approximately $\Delta t =4$ps. 
(d) : Close up of the snapshot (c), 
in which a step structure with a six-membered ring appears. 
}
\label{fig-pandey-step-short}
\end{figure}%

\subsection{Stable cleavage mode on $(111)$ surface with $(2 \! \times \! 1)$  reconstruction}

Our simulation results reproduce 
the stable (experimental) cleavage mode 
with the $\pi$-bonded $(111)$-$(2 \! \times \! 1)$ surface  
that was proposed first by Pandey \cite{PANDEY} and 
is currently well established in both experiment and theory. 
\cite{NOTE-HOSHI2004A} 
Since a surface atom on the {\it ideal} $(111)$ surface has one dangling-bond electron,
the pairing mechanism between the nearest neighbor dangling-bond electrons 
is the motive force of the surface reconstruction. 
Figure \ref{fig-hane-pan-fin} shows the actual reconstruction process
with quantum mechanical analysis of the wavefunction.  
The resultant surface contains a pair of 
five- and seven-membered rings 
as the unit of the $\pi$-bonded $(2 \! \times \! 1)$ structure. 
The $\pi$-bonding appears between the $B$ and $D$ sites
in Fig.~\ref{fig-hane-pan-fin}(e), 
as indicated by an oval. 

\subsection{Step structure and bias-dependent STM image}

In our results, 
step formation were frequently observed on 
the $(111)$-$(2 \!  \times \! 1)$ surface.  
The frequently observed step appears in 
Fig.~\ref{fig-pandey-step-short}(d) and has
a six-membered ring at the step edge.
The formation process was investigated 
with quantum mechanical (electronic) freedom
and we found the mechanism that 
explains why this kind of step appears 
so frequently. \cite{HOSHI2004A} 
Among STM experiments on cleaved surface, on the other hand, 
several step structures were observed  but 
their explicit atomic structures have not been settled. 
\cite{FEENSTRA87,TOKUMOTO,MERA92}
Since our results were independent from the experiments, 
we propose the present step structure as a hopeful candidate.

So as to compare with STM experiments, 
the local density of states (LDOS) was 
calculated for the surface atoms,
the $A,B,C,D$ sites in Fig.~\ref{fig-pandey-step-short}(d), 
using the Krylov subspace method for Green function.  
\cite{TAKAYAMA2004B,TAKAYAMA2004B-CODE}
In result,  
the upper (vacuum side) atom ($A$ or $C$ site)
has an {\it occupied} surface state 
and the lower (bulk side) atom ($B$ or $D$ site) 
has an {\it unoccupied} surface state. 
The present LDOS result 
corresponds to a bias-dependent STM image.
Actually, such a bias-dependent STM image is observed experimentally 
\cite{FEENSTRA86} on the flat (non-stepped) region, 
the $A$ and $B$ sites in Fig.~\ref{fig-pandey-step-short}(d). 
Therefore, we concluded that such a bias-dependent STM image 
will be observed also in the stepped region, 
the $C$ and $D$ sites in Fig.~\ref{fig-pandey-step-short}(d). 
More quantitative discussions will appear elsewhere.

\section{Summary}

Practical methods of large-scale electronic structure 
were constructed and their foundation is to calculate 
 (one-body) density matrix or Green function,
instead of eigenstates. 
In the application of cleavage process of silicon,  
nanoscale defects, step and bending, were observed
as well as elementary surface reconstruction processes.
All the results are consistent to experiments
qualitatively and quantitatively.  
Moreover,  our simulation results predict 
a practical step structure  
with the bias dependency of its STM image.
 
Since the present method for large systems is a general quantum mechanical calculation,
it has wide applications, not specific to silicon or cleavage.
Processes of other nanoscale or 10-nm-scale systems  
are possible targets in future study.

\section*{Acknowledgements}

Numerical calculation was partly carried out 
using the facilities of the Japan Atomic Energy Research Institute, 
the Research Center for Computational Science, Okazaki and
the Institute for Solid State Physics, University of Tokyo.


\begin{thebibliography}{}

\bibitem{CP}
R. Car and M. Parrinello, 
Phys. Rev. Lett. {\bf 55}, 2471 (1985). 


\bibitem{KOHN96}
 W. Kohn, Phys. Rev. Lett. {\bf 76}, 3168 (1996). 

\bibitem{REVIEW-ON}
For reviews;
 G. Galli, Phys. Status Solidi B{\bf 217}, 231 (2000); 
 S. Y. Wu and C. S. Jayamthi, Phys. Rep. {\bf 358}, 1 (2002).

 \bibitem{HOSHI2004A}
T. Hoshi, Y. Iguchi and T. Fujiwara, 
Phys. Rev. B{\bf 72}, 075323 (2005).

 
 \bibitem{HOSHI2000A}
T. Hoshi and T. Fujiwara,
J. Phys. Soc. Jpn. {\bf 69}, 3773 (2000).


\bibitem{HOSHI2001A}
T. Hoshi and T. Fujiwara,
Surf. Sci. {\bf 493}, 659 (2001).

\bibitem{HOSHI2003A}
T. Hoshi and T. Fujiwara,
J. Phys. Soc. Jpn. {\bf 72}, 2429 (2003).

\bibitem{GESHI}
M. Geshi, T. Hoshi, and T. Fujiwara,
J. Phys. Soc. Jpn. {\bf 72}, 2880 (2003).

\bibitem{TAKAYAMA2004A}
R. Takayama, T.Hoshi, and T.Fujiwara,
J. Phys. Soc. Jpn. {\bf 73}, 1519 (2004).


\bibitem{TAKAYAMA2004B}
R. Takayama, T. Hoshi, T. Sogabe, S.-L. Zhang and T. Fujiwara, 
preprint (cond-mat/0503394).

\bibitem{TAKAYAMA2004B-CODE}
The program code for calculating Green function is available 
at http://act.jst.go.jp. 

\bibitem{KWON} 
 I.~Kwon, R.~Biswas, C.~Z.~Wang, K.~M.~Ho, and C.~M.~Soukoulis, 
 Phys. Rev. B {\bf 49}, 7242 (1994). 


\bibitem{KRYLOV}
Krylov subspace is a general mathematical concept
in linear algebra. 
As a recent textbook,
H. A. van der Vorst,
{\it Iterative Krylov methods for large linear systems},
Cambridge University Press (2003).

\bibitem{LMTO}
O. K. Andersen and O. Jepsen, Phys. Rev. Lett. 53, 2571
(1984).

 \bibitem{NOTE-HOSHI2004A-2}
See Refs. (37-42) of Ref. \cite{HOSHI2004A}

 \bibitem{MOTT}
 N. F. Mott, Engineering 165, 16 (1948).
 
 \bibitem{BRITTLE}
As a textbook,  B. Lawn,  
{\it Fracture of brittle solids}, 2nd ed., 
Cambridge University Press (1993).

\bibitem{SPENCE93}
Y. M. Huang, J. C. H. Spence, O. F. Sankey, and G. B. Adams,
Surf. Sci. {\bf 256}, 344 (1991); 
J.~C.~H. Spence, Y.~M. Huang, and O. Sankey, 
Acta Metall. Mater. {\bf 41}, 2815  (1993).

\bibitem{GUMBSCH}
R. P\'erez and P. Gumbsch,
Phys. Rev. Lett. {\bf 84}, 5347 (2000).

\bibitem{HAUCH1999}
J. A. Hauch, D. Holland, M. P. Marder
and H. L. Swinney, Phys. Rev. Lett. 82, 3823 (1999).


\bibitem{PANDEY}
K. C. Pandey, Phys. Rev. Lett. {\bf 47}, 1913 (1981).


 \bibitem{NOTE-HOSHI2004A}
See References in Ref. \cite{HOSHI2004A}


\bibitem{FEENSTRA87}
R. M. Feenstra and J. A. Stroscio,
Phys. Rev. Lett. {\bf 59},  2173  (1987).

\bibitem{TOKUMOTO}
H. Tokumoto, S. Wakiyama, K. Miki, H. Murakami,
S. Okayama, and K. Kajimura,
J. Vac. Sci. \& Technol. B {\bf 9} 695 (1991);
T. Komeda, S. Gwo, and H. Tokumoto,
Jpn. J. Appl. Phys. {\bf 35}, 3724 (1996).

\bibitem{MERA92}
Y. Mera, T. Hashizume, K. Maeda, and T. Sakurai,
Ultramicroscopy {\bf 42-44}, 915 (1992).


\bibitem{FEENSTRA86}
R. M. Feenstra, W. A. Thompson and A. P. Fein
Phys. Rev. Lett. {\bf 56},  608  (1986);
J. A. Stroscio, R. M. Feenstra and A. P. Fein
Phys. Rev. Lett. {\bf 57},  2579  (1986).


\end{thebibliography}
\end{document}